\documentclass[num-refs]{wiley-article}
\usepackage{graphicx}
\usepackage[hidelinks]{hyperref}
\usepackage{subfigure}
\usepackage{algorithm}
\usepackage{algorithmicx}
\usepackage{algpseudocode}
\usepackage{booktabs}

\usepackage[per-mode=symbol, binary-units, detect-all]{siunitx} 
\DeclareMathOperator{\bitwiseand}{\land}
\DeclareMathOperator{\bitwiseor}{\lor}
\DeclareMathOperator{\bitwisexor}{\oplus}

\DeclareMathOperator{\maj}{maj}

\usepackage{listings}
\lstdefinestyle{customc}{%
  frame=ltbr,framesep=4pt,
  xleftmargin=4pt,
  xrightmargin=4pt,
  belowcaptionskip=1\baselineskip,
  breaklines=true,
  language=C,
  showstringspaces=false,
  basicstyle=\scriptsize\ttfamily,
  keywordstyle=\bfseries\color{blue},
  keywordstyle=[2]\bfseries\color{green!60!black}, 
  keywordstyle=[3]\fontfamily{lmtt}\fontseries{b}\selectfont\color{blue!60!black},           
  numberstyle=\tiny,
  commentstyle=\itshape\color{purple!60!black},
  identifierstyle=\bfseries\color{black},
  stringstyle=\color{orange},
  keywords={for,if,while}, 
  morekeywords=[2]{size_t,void,int,uintw_t,uint64_t,uint32_t,__m256i,__m128i,simd8,uint8_t,UINT64_C},
  morekeywords=[3]{pospop_w},
  escapeinside={(*@}{@*)}
}

\usepackage{xcolor}

\begin{document}
\newcommand{\pospopcnt}{\textsf{pospopcnt}}

\newcommand{\mytitle}{Faster Positional-Population Counts for AVX2, AVX-512, and ASIMD}
\title{\mytitle{}}
\author[1]{Robert Clausecker}
\author[2]{Daniel Lemire}
\author[1]{Florian Schintke}
\runningauthor{R. Clausecker, D. Lemire, F. Schintke}
\affil[1]{Distributed Algorithms, Zuse Institute Berlin}
\affil[2]{DOT-Lab Research Center, Université du Québec ({TÉLUQ})}
\corraddress{Robert Clausecker, Zuse Institute Berlin, Takustra\ss e 7, 14195~Berlin, GERMANY}
\corremail{clausecker@zib.de}

\maketitle

\begin{abstract}
The positional population count operation \pospopcnt\ counts for an array
of $w$-bit words how often each of the $w$~bits was set.  Various applications
in bioinformatics, database engineering, and digital processing exist.

Building on earlier work by Klarqvist~et~al., we show how positional population
counts can be rapidly computed using SIMD~techniques with good performance from
the first byte, approaching memory-bound speeds for input arrays of as little
as~\SI{4}{\kibi\byte}.  Improvements include an improved algorithm structure,
better handling of unaligned and very short arrays, as well as faster
bit-parallel accumulation of intermediate results.

We provide a generic algorithm description as well as implementations for
various SIMD instruction set extensions, including Intel AVX2, AVX-512, and
ARM ASIMD, and discuss the adaption of our algorithm to other platforms.
\end{abstract}

\section{Introduction}
Low-cardinality categorical variables are often represented using one-hot encoding~\cite{lippert2013exhaustive,mittag2015influence}: each categorical value is associated with a  bit within a $w$-bit word. For example, given the variable \texttt{age}, one might have four distinct age categories for the ages between \texttt{0-20}, \texttt{21-35}, \texttt{36-65}, \texttt{66-120} years. 
We may represent each category value using a 4-bit word: 
\texttt{000\underline{1}}, \texttt{00\underline{1}0}, 
\texttt{0\underline{1}00} and \texttt{\underline{1}000}.
From these words, we would like to compute as quickly as possible the histogram: the number of occurrences of each value.

\begin{figure}
    \centering
    \includegraphics[width=0.3\linewidth]{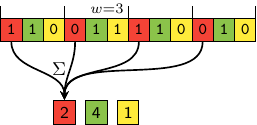}
    \caption{Positional population count with $w=3$}
    \label{fig:post3}
\end{figure}

For such purposes, Klarqvist et~al.~\cite{Klarqvist2019} introduced the \emph{positional population count}. The conventional population count is merely the sum of the bit values (0 and 1) in a stream of bits. It is an important operation in databases and cryptography: most commodity processors (ARM, x64) have dedicated instructions to accelerate the computation of the conventional population count.
When we compute the \emph{positional} population count, we  view a stream of bits as $w$~interleaved streams for some integer parameter $w$: we sum the bit values at position 0, $w$, $2w$, \ldots;
we sum the bit values at 1, $w+1$, $2w+1$, \ldots; \ldots;
we sum the bit values at $w-1$, $2w-1$, $3w-1$, \ldots\ So, the positional population count provides $w$~distinct sums. For $w>1$, positional population count generalizes the conventional population count. See Fig.~\ref{fig:post3}.

Given a database of one-hot encoded values, the positional population count might accelerate group-by queries (e.\,g. \texttt{SELECT COUNT(*), country FROM table GROUP BY country}). Furthermore, the positional population count is not limited to one-hot encoded data: it computes histograms over arbitrary bits which could be useful for various statistical tests.
Other uses are found in the construction of \emph{wavelet trees}\cite{Dinklage2021}, and in approximate pattern matching of DNA sequences\cite{Bingmann2019}, an application where an early form of our implementation is already in use\cite{KMCP22}.

To compute the positional population count, we might  proceed by accessing each bit value in sequence, as in Fig.~\ref{fig:pseudocpp}. It is likely that such code requires several processor instructions per input bit.
In contrast, our objective is to spend few instructions per input word. 

\begin{figure}
    \centering
\lstset{escapechar=@,style=customc}

\begin{lstlisting}[identifierstyle={\bfseries\color{violet!90!white}}]
void pospop_w(int counts[w], uintw_t words[], size_t n) {
        for (size_t i = 0; i < n; i++)
                for (size_t j = 0; j < w; j++)
                        counts[j] += words[i] >> j & 1
}
\end{lstlisting}
\caption{Pseudo-C function to compute the width-$w$ positional population count\label{fig:pseudocpp}}
\end{figure}

Conventional population count can be seen as a `horizontal' operation as the set bits across a word or array are counted---independent of their position.
In contrast, positional population count has a `vertical' component as only every $w$-th bit is to be accumulated in the same counter, which should suit the calculation well for SIMD instructions of modern microprocessors.
Modern processors have wide registers, spanning up to 512~bits in recent Intel processors, with accompanying AVX-512 single-instruction-multiple-data (SIMD) instructions~\cite{intel-avx2-avx512}. These instructions can execute the same operation on multiple values at once. For sequences of machine words in contiguous memory, SIMD instructions help us compute the conventional population count at gigabytes per second~\cite{mula2017faster}.  Using similar techniques, Klarqvist et~al.~\cite{Klarqvist2019} showed that we can achieve gigabytes per second when computing the positional population count by using SIMD instructions as well.
Their methods, unfortunately, need inputs exceeding a few kilobytes in size to observe notable speed improvements compared to scalar code.

Our main contribution is a refined algorithm for the positional population
count operation, building on Klarqvist~et~al.'s work~\cite{Klarqvist2019}.
In particular,
\begin{itemize}
 \item we show how Harley-Seal algorithm scheme can be improved using a simplified
 CSA~network for the initial iteration (see \S~\ref{sec:csareduction}),

 \item we provide effective approaches to deal with inputs that are not aligned
 to the vector length (see \S~\ref{sec:headprocessing}) as well as very short
 inputs (see \S~\ref{sec:short-arrays-scalar-tail}), yielding good performance
 from the first byte,

 \item we demonstrate how the CSA-summed bit vectors generated in the
 main loop of Klarqvist~et~al.'s algorithm can be transposed and rapidly
 added to the accumulator vectors in a bit-parallel manner (see \S~\ref{sec:intermedaccum}
 and \S~\ref{sec:final-accumulation}),

 \item we implement our algorithm for different SIMD instruction set extensions
 including Intel AVX2, AVX-512 and ARM ASIMD and compare its performance with
 the Klarqvist~et~al.\ algorithm for a variety of input sizes, showing that
 its execution is memory bound with inputs as small as \SI{4}{\kibi\byte}
 (see \S~\ref{sec:evaluation}).
\end{itemize}

\section{SIMD Instruction Set Extensions}
Our algorithms are designed to be implemented using \emph{SIMD
instruction set extensions}.
SIMD (single instruction multiple data)~\cite{Flynn1972}
extensions generally provide additional CPU registers holding \emph{vectors} of data
items and instructions that operate on the items of their vector operands in parallel
(\emph{vertical instructions}).
These include instructions such as vector loads and stores, addition and subtraction,
as well as bitwise operations.
A number of instructions instead perform arithmetic or data movement between the
elements of one vector (\emph{horizontal instructions}).
These generally include
\emph{reductions} (such as instructions to sum the elements of a vector) and
\emph{shuffles} (changing the order of vector elements, either according to a fixed
scheme or by a user-defined permutation).
In recent SIMD extensions such as AVX-512 and SVE, SIMD registers may be complemented
by \emph{predicate mask} registers, allowing the programmer to decide which vector
elements an instruction is to affect.

We consider three families of SIMD instruction set extensions.
The \emph{AVX} (advanced vector extensions) family comprising AVX and AVX2,
extending the Intel~64 instruction-set architecture\footnote{the 64-bit variant of the
IA-32 (x86, i386) architecture also known as x86-64, x64, amd64, IA-32e, and EM64T},
the \emph{AVX-512} family comprising a variety of extension sets, of which we use the
\textbf{F}~(\emph{foundation}) and \textbf{BW}~(\emph{byte and word instructions})
sets, also extending Intel~64,
and the \emph{ASIMD} (advanced SIMD) extension to AArch64.
We also briefly mention other SIMD extensions, but those three are the focus of
this publication.

\subsection{AVX2}
An upgrade to the earlier SSE~family of instruction set extensions, AVX2
provides the programmer with sixteen SIMD registers of 256~bits that can be subdivided
into elements of 8, 16, 32, or 64~bits.
A comprehensive set of the usual vertical integer and floating point operations is
provided, with the notable omission of 8~bit shift instructions.
Horizontal reductions and shuffles complement the set, but no full arbitrary
32-byte (or sixteen 16-bit-word) permutation instructions are provided, necessitating
careful algorithm design to work around this limitation.

AVX2 registers can be thought of as being divided into two 128-bit (or 16-byte) lanes,
with most instructions performing identical and independent operation on both lanes.
This simplifies the transition from the earlier 128-bit wide SSE family of instruction
set extensions, which can be seen as operating like AVX2, but with only one lane per
register.
A small number of \emph{cross-lane} instructions break with this pattern and
provide means to exchange data across the lanes of an AVX2 register, but at an
increased latency.
For example, on Intel \emph{Icelake} processors, the in-lane \texttt{vpshufb}
(packed shuffle bytes) instruction permutes bytes within 128-bit lanes in one cycle
at a throughput of two per cycle.
Meanwhile, the cross-lane \texttt{vpermd} (packed permute doublewords) instruction
permutes 32-bit words within the entire register in three cycles at a throughput of
one per cycle.
This design feature encourages the programmer to find algorithmic approaches that
minimise the number of cross-lane shuffles.

\subsection{AVX-512}
AVX-512 extends AVX2 to 512-bit registers while keeping the overall architecture and
design of~AVX2.
The number of SIMD registers is doubled to~32.
A number of new instructions are provided, including additional permutation
instructions and \texttt{vpternlogd} (packed ternary logic doubleword), an
instruction to compute any three-input one-output bitwise operation based on a
truth table supplied as an immediate operand.
For example, the bitwise ternary operator $a=a\mathrel?b:c$ can be computed using
immediate operand~\texttt{0xca} as \texttt{vpternlogd a, b, c, 0xca}, representing
the following truth table:

\begin{center}
\begingroup
\setlength{\tabcolsep}{2.5pt}
\begin{tabular}{r|llllllll}
$a$&1&1&1&1\,&0&0&0&0\\
$b$&1&1&0&0  &1&1&0&0\\
$c$&1&0&1&0  &1&0&1&0\\
\hline
\texttt{0xca} $\widehat{=}$ $a\mathrel?b:c$&1&1&0&0&1&0&1&0\\
\end{tabular}
\endgroup
\end{center}

As another notable new feature, \emph{mask registers} can be used to decide
which vector elements are affected by a SIMD instruction.
Other elements remain unchanged (\emph{merge masking}) or may be cleared
(\emph{zero masking}), depending on instruction and masking mode.
When applied to instructions with memory source operands, no faults occur from
masked out elements (\emph{fault suppression}).  When applied to stores, merge masking
can be used to store possibly discontiguous data without affecting adjacent data.

On the hardware side, Intel implements AVX-512 on the Skylake microarchitecture using
the same execution ports 0, 1, and~5 as used for AVX2.
For AVX2 instructions, these ports are used at a width of 256~bits,
permitting up to three AVX2 instructions to be dispatched per cycle.
With AVX-512, port~5 is extended to 512~bits.
Instructions operating on 512-bit vectors are either executed on port~5, or the SIMD
resources of ports 0 and~1 are
bundled to execute one instruction together, for a total of up to two instructions per
cycle.\footnote{Such an instruction formally runs on port~0; port~1 can execute another
non-SIMD instruction in the same cycle.}
Thus while 256-bit vectors can be processed at up to 6~lanes of 128~bits each per cycle,
512-bit~vectors raise this number to only 8~lanes for only a 33\,\%~performance increase,
despite the doubled vector length.
At the same time, on some microarchitectures, CPU frequency is reduced due to
\emph{thermal licensing}\cite{Downs2020, Lemire2018}
when executing SIMD instructions at full 512~bits vector length.

Despite these limitations, use of 512-bit vectors poses many advantages in this
application:
less instruction-level parallelism (ILP) is needed to process the same amount of data,
permutations over wider vectors reduce the total number of transposition steps needed,
register pressure is reduced, allowing more flexibility with instruction ordering,
and the presence of \texttt{vpternlogd} significantly reduces the number of steps per
full adder (see \S~\ref{sec:csaimpl}).

\subsection{ASIMD}
The \emph{Advanced SIMD} (ASIMD) instruction set extension provides
SIMD instructions for the AArch64 architecture.
The feature set is comparable to that of AVX2, but with 32~registers of only 16~bytes.
At the same time, powerful bulk load/store instructions transfer up to 64~bytes at a
time.
ASIMD is characterised by a more orthogonal set of integer instructions, with almost all
instructions being available at all data sizes.
Arithmetic instructions often support sign and zero extension as a side effect of their
operation, reducing the need for explicit shuffles.
A flexible generic 4-input byte permutation instruction~\texttt{tbl} is provided, allowing
the programmer to pick any 16~bytes out of a 64-byte look-up table formed from four consecutive
registers.

Performance of ASIMD instructions varies considerably across implementations, with low-end
ARM processors some times even seeing a disadvantage over scalar code, while high-end
out of order designs provide performance that is on par or even exceeding that of AVX2 on
an Intel chip clocked at the same frequency.
Another big difference is found in the performance of permutation instructions: while these
perform reasonably well on high-end cores, many cores do not cope well with wide permutations
and it is often useful to seek other approaches where possible.

This diversity in implementations renders the design of SIMD algorithms with good performance
across CPU designs a very challenging task.
Neverthless we believe that we found a good compromise in the ASIMD~implementation of our
algorithm.

\section{Background} \label{sec:background}

The key to the efficient computation of positional population counts as well as
regular population count lies in \emph{carry-save adder (CSA) networks}~\cite{csa-networks}.
These are built by combining \emph{bit-parallel full adders} into networks, that can then
be used to compute the population of input vectors in chunks.

\subsection{Bit-parallel full adders}
A full-adder (FA) circuit takes three one bit inputs and produces a two bit
output representing the number of input bits that have been set.
The lower output bit~$\Sigma$ is called the \emph{sum},
the upper output bit~$C$ is the \emph{carry}.
This can be seen as a compression of the population of three input bits
weighted~$(1,1,1)$ into two output bits weighted~$(1,2)$.

Using standard bitwise instructions, we can treat SIMD registers as vectors of
bits and simulate a full adder circuit on a whole vector worth of bits in parallel,
taking three vectors of input and giving vectors of sum and carry as output.
For example, at a vector length of $r=4$, given the three vectors
$a=\texttt{1001}$, $b=\texttt{1001}$, and~$c=\texttt{0101}$,
we can simulate a full adder to produce $C=\texttt{1001}$ and $\Sigma=\texttt{0101}$.
We effectively simulate four full adders, one for each element in our 4-bit vectors.

This also highlights the difference to a normal addition routine: while an
addition operation~$\Sigma=a+b$ is a \emph{horizontal} operation in which a
multi-bit number~$a$ is added to a multi-bit number~$b$, the bit-parallel full
adder routine is a vertical SIMD operation~$(C,\Sigma)=\mathit{FA}(a,b,c)$ in which each
bit is processed independently.

\subsubsection{Implementation details} \label{sec:csaimpl}
The bit-parallel full-adder circuit can be realised in a variety of ways.
On all architectures, the classic five gate full adder circuit (Eq.~\ref{eqn:fivegate}),
comprising two half adders and an \emph{or} operation to combine their carry-outs
can be used. This strategy and its variants require 5~operation steps with a
critical path length of 3~steps.
It is used to implement the full-adder operation on architectures that only provide
basic two-input logic operations such as x86 with MMX, SSE, and AVX2, ARM with SVE,
as well as when implementing bit-parallel full adders in high-level languages.

\begin{minipage}{.22\textwidth}
\begin{equation}\label{eqn:fivegate}
\begin{alignedat}{3}
\Sigma_1&=~a &\bitwisexor &~b\\[-1mm]
C_1&=~a &\bitwiseand &~b\\[-1mm]
\Sigma&=\Sigma_1 &\bitwisexor &~c\\[-1mm]
C&=C_1 &\bitwiseor&~(\Sigma_1\bitwiseand c)\\
\end{alignedat}
\end{equation}
\end{minipage}
\hfill\hfill
\begin{minipage}{.70\textwidth}\vspace*{4pt}
\begin{lstlisting}[style=customc,language={[Motorola68k]Assembler},morecomment={[l]{\#}},morecomment={[l]{;}},morekeywords={vpand,vpxor,vpor,bit,vmovdqa64,vpternlogd}]
; AVX2: sum bit vectors ymm0, ymm1, ymm2 into ymm1:ymm0
vpand ymm3, ymm0, ymm1  ; C1  = a  & b
vpxor ymm0, ymm0, ymm1  ; S1  = a  ^ b
vpand ymm1, ymm0, ymm2  ; tmp = S1 & c
vpxor ymm0, ymm0, ymm2  ; S   = S1 ^ c
vpor  ymm1, ymm3, ymm1  ; C   = C1 | tmp
\end{lstlisting}
\end{minipage}

\noindent
Some SIMD units provide a `mux' instruction implementing the ternary operator
$a\mathbin?b\mathbin:c$ for each bit.
In this case, a more efficient implementation (Eq.~\ref{eqn:mux}) using just
3~operations with a 2~operation critical path length can be used.
The sum~$\Sigma$ is again computed by two exclusive-or operations
while the carry out~$C$ is taken using the mux instruction.
This variant is used on SIMD units such as ARM ASIMD (using \texttt{bsl},
\texttt{bit}, or~\texttt{bif}) and POWER VMX (using \texttt{vsel}).

\begin{minipage}{.22\textwidth}
\begin{equation}\label{eqn:mux}
\begin{alignedat}{3}
\Sigma_1&=a & \bitwisexor & ~b\\[-1mm]
\Sigma&=\Sigma_1 & \bitwisexor & ~c\\[-1mm]
C&=\Sigma_1 & \mathbin? & ~c\mathbin:b\\
\end{alignedat}
\end{equation}
\end{minipage}
\hfill\hfill
\begin{minipage}{.70\textwidth}\vspace*{4pt}
\begin{lstlisting}[style=customc,language={[Motorola68k]Assembler},morecomment={[l]{\#}},morecomment={[l]{//}},morekeywords={vpand,vpxor,vpor,bit,vmovdqa64,vpternlogd}]
// AArch64 ASIMD: sum bit vectors v0, v1, v2 into v1:v0
eor   v3.16b, v0.16b, v1.16b
eor   v0.16b, v3.16b, v2.16b  // S = v0 = v0 ^ v1 ^ v2
bit   v1.16b, v2.16b, v3.16b  // C = v1 = v0^v1 ? v2 : v1
\end{lstlisting}
\end{minipage}

\noindent
Finally, AVX-512 provides the \texttt{vpternlogd} instruction to perform
any three-input single-output bitwise operation based on the given truth table
(see~\cite[Volume 2c, Section 5.1]{intel-avx2-avx512}).  This allows us to
compute $\Sigma$ and~$C$ directly from the inputs: the sum as the parity, the
carry as the majority (i.\,e. whether more bits are set than not).
These two operations are represented by the truth tables \texttt{0x96} and
\texttt{0xe8} for parity and majority respectively.

As \texttt{vpternlogd} is destructive (i.\,e. overwrites one of its operands),
we have to use an additional data move to preserve all three input registers
through the first of the two \texttt{vpternlogd}~instructions.
All implementations of AVX-512 available on the market as of the writing of
this article implement such data moves as zero-latency register renames
(i.\,e. they are effectively free).
Hence, even though three instructions are involved in this implementation,
we count this as 2~operation steps with a critical path length of 1~step.

\begin{minipage}{.20\textwidth}
\begin{equation}\label{eqn:lut}
\begin{split}
\Sigma&=a\bitwisexor b\bitwisexor c\\[-1mm]
C&=\maj(a,b,c)\\
\end{split}
\end{equation}
\hfill\hfill
\end{minipage}
\hfill\hfill
\begin{minipage}{.73\textwidth}\vspace*{4pt}
\begin{lstlisting}[style=customc,language={[Motorola68k]Assembler},morecomment={[l]{\#}},morecomment={[l]{;}},morekeywords={vpand,vpxor,vpor,bit,vmovdqa64,vpternlogd}]
; AVX-512: sum bit vectors zmm0, zmm1, zmm2 into zmm1:zmm0
vmovdqa64  zmm3, zmm0               ; zmm3 = zmm0
vpternlogd zmm0, zmm1, zmm2, 0x96   ; S = a ^ b ^ c
vpternlogd zmm1, zmm3, zmm2, 0xe8   ; C = maj(a, b, c)
\end{lstlisting}
\end{minipage}

\begin{figure}
\centering
\subfigure[CSA$_{15}$ network for the initial 15 registers]{
\includegraphics[width=0.46\textwidth]{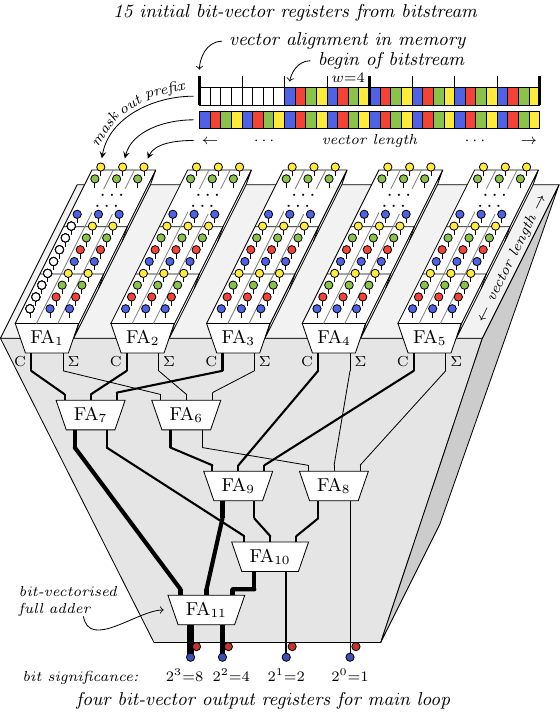}
\label{csa:15to4}}\quad
\subfigure[CSA$_{16+4}$ network for the main loop (bit parallelism akin to (a), not shown here)]{
\includegraphics[width=0.46\textwidth]{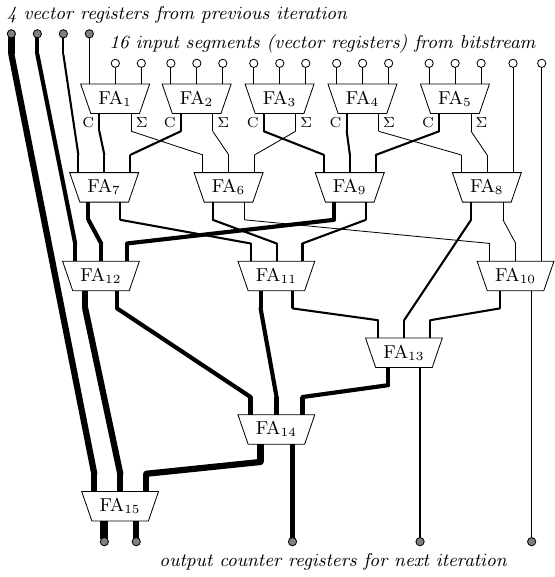}
\label{csa:16to5}}
\caption{Bit-parallel carry-save adder (CSA) networks
\label{fig:csa-networks}}
\end{figure}

\subsection{Carry-Save Adder (CSA) Networks} \label{sec:csa}
Bit-parallel full-adders provide us with a building block to compress,
i.\,e. count, the population of three bits into two.
Counting more bits is achieved by repeatedly compressing bits of the same
weight with full adders until this is no longer possible.
For example, seven input bits weighted~$(1,1,1,1,1,1,1)$ can be compressed into
three output bits with weights~$(1,2,4)$.  15~input bits weighted~$(1,1,\ldots,1)$
can be compressed into four output bits with weights~$(1,2,4,8)$.

To visualise this process, we draw these sequences of full adders as
\emph{CSA networks} (see Fig.~\ref{fig:csa-networks}).
This visualisation highlights that a great deal of full adders can be
evaluated in parallel, improving the \emph{instructions per cycle (IPC)} on out-of-order processors.  For example,
the CSA$_{15}$~network in Fig.~\ref{csa:15to4} performs 11~(vectorised) full-adder operations with a
critical path of only 5~FA operations.

\subsection{Counting bits with CSA Networks}\label{sec:pospopcount}
The use of CSA~networks to count the bits in arrays was first proposed
by Harley and Seal in 1996 and later popularised by
Warren\cite{Warren2013}.  By first reducing the $2^k$~words of an array to
$k+1$~accumulators $a_{2^{k+1}},\ldots,a_2,a_1$
of place-value $2^{k+1},\ldots,2,1$,
costly population count steps can be reduced to comparably cheap bitwise logic
and only few final population count steps to sum up the bit-parallel accumulators.

In practice, a fixed block size of 8 or~16 words is chosen and after each chunk of
that many words, the population of the most-significant accumulator resulting from the
CSA~network is taken and added to the population count, while the other accumulators are
carried over into the next iteration.  This reduces the algorithm to one actual population
count per chunk of input; the population of the remaining accumulators resulting from the
CSA~network needs only be computed once after the final
iteration (see Alg.~\ref{alg:harleyseal}).

\begin{algorithm}
\caption{The Harley-Seal algorithm\cite{Warren2013}} \label{alg:harleyseal}
\begin{algorithmic}
\Require{\textrm{$A$ is an array of $n$ blocks of $2^k$~words each}}
\State $(a_{2^{k-1}}, \ldots, a_2, a_1)\leftarrow(0,\ldots,0)$
\State $c\leftarrow0$
\For{$i\leftarrow0,2^k,\ldots,(n-1)2^k$}
\State $\triangleright$ \textrm{combine $2^k$~words of input with accumulators $(a_{2^{k-1}},\ldots,a_2,a_1)$}
\State $a_{2^k},(a_{2^{k-1}}\ldots,a_2,a_1)\leftarrow\textrm{CSA}(a_{2^{k-1}},\ldots,a_2,a_1;\;A[i],A[i+1],\ldots,A[i+2^k-1])$
\State $c\leftarrow c+2^k\,\textrm{popcount}(a_{2^k})$ 
\EndFor
\State $c\leftarrow c+\sum_{i=0}^{k-1}2^i\,\textrm{popcount}(a_{2^i})$ 
\State \Return $c$
\end{algorithmic}
\end{algorithm}

This idea was combined by Mu\l a et~al.\ with a novel table-based population count
routine (see Alg.~\ref{alg:klarqvist}) to yield the currently fastest known
algorithm for counting the bits of arrays.\cite{mula2017faster}
Later, the same Harley and Seal algorithm structure was used by Klarqvist
et~al.\cite{Klarqvist2019} to count bits in an array grouped by their places in
each word in a vectorised fashion, yielding a fast \emph{positional population
count} algorithm for the first time.

\begin{algorithm}
\caption{Sketch of the Klarqvist~et al.~algorithm without vectorization\cite{Klarqvist2019}} \label{alg:klarqvist}
\begin{algorithmic}
\Require{\textrm{$A$ is an array of $n$ blocks of $2^k$ words of $w$~bits each}}
\State $(a_{2^{k-1}}, \ldots, a_2, a_1)\leftarrow(0,\ldots,0)$
\State $c[0,1,\ldots,w-1]\leftarrow0$ 
\For{$i\leftarrow0,2^k,\ldots,(n-1)2^k$}
\State $\triangleright$ \textrm{combine $2^k$~words of input with accumulators $(a_{2^{k-1}},\ldots,a_2,a_1)$}
\State $a_{2^k},(a_{2^{k-1}},\ldots,a_2,a_1)\leftarrow\textrm{CSA}(a_{2^{k-1}},\ldots,a_2,a_1;\;A[i],A[i+1],\ldots,A[i+2^k-1])$
\For{$j\leftarrow0,1,\ldots,w-1$}
\State $c[j]\leftarrow c[j]+2^k((a_{2^k}\gg j)\bitwiseand 1)$ 
\EndFor
\EndFor
\For{$j\leftarrow0,1,\ldots,w-1$}
\State $c[j]\leftarrow c[j]+\sum_{i=0}^{k-1}2^i((a_{2^i}\gg j)\bitwiseand 1)$ 
\EndFor
\State \Return $c$
\end{algorithmic}
\end{algorithm}

However, while vectorized, the Klarqvist~et al.\ procedure
is at its heart still based on a scalar procedure.  After the CSA~step of the
main loop, the accumulators need to be \emph{accumulated} into the counters.  Each
bit of the most-significant accumulator~$a_{2^k}$ is individually
added to the corresponding counter, leading to a number of steps proportional to
the word width~$w$ by which we want to group bits.  The same cumbersome procedure
takes place at the end to sum the remaining accumulators into the counter array.
While vectorization is employed to process multiple words at the same time,
there is clearly room for improvement.

We aim to improve on this result by providing fully
bit-parallel accumulation procedures that reduce the amount of steps needed
to~$\mathcal O(\log w)$.  We also show how the Harley-Seal algorithm scheme can
be improved for lower startup cost and provide methods to deal with unaligned
input, very short input (i.\,e. less than one iteration of the main loop), and
the tail that remains after no more iterations of the main loop can be executed.

\section{Fast SIMD-vectorized positional-population counts}
Like the Klarqvist et al. algorithm (see \S~\ref{sec:pospopcount} and Alg.~\ref{alg:klarqvist}), our algorithm takes an array~$A$, computes its positional
population count with respect to some word size~$w$, a power of two, and adds the population
to an array~$C$ of $w$~counters.  It is expected that $A$~is aligned to a multiple
of~$w$.

The algorithm internally computes the population count with respect to some maximum
word size~$w_{\max}\ge w$ in vectors of some vector size\footnote{The vector size~$r$ is
usually the CPU's native vector size, so $r=128$ for SSE, ASIMD, and VMX, $r=256$ for
AVX2, and $r=512$ for AVX-512.}~$r\ge w_{\max}$, both
of which must be powers of~2, too.
For each desired word size~$w$, an accumulation function~$f_w(C, c)$ must be provided that
reduces the internal counter vector~$c$ to $w$~elements and adds the reduced counters to
the counter array~$C$.  This permits use of the same code for different word sizes.  In the
code developed for this paper, $w_{\max}=64$ was chosen for all implementations.

If the input is less than $15r$~bytes in total, we proceed
with the special processing from \S~\ref{sec:short-arrays-scalar-tail}.  Otherwise:

\begin{enumerate}
\item Optionally, clear the counter array~$C$.\footnote{If this is omitted, counts of
the current array are added to the counts already in~$C$.  This may be desirable for
streamed input or non-contiguous memory regions to be counted.}
\item Align to a multiple of~$r$ and perform a load of one vector $v_0$ from the
aligned address.  Clear all bytes in $v_0$ that precede the beginning of the array (\S~\ref{sec:headprocessing}).
\item Load 14~more vectors $v_1, v_2, \ldots, v_{14}$ from the input array and
reduce $v_0, v_1, \ldots, v_{14}$ into accumulators~$(a_8, a_4, a_2, a_1)$ using a
CSA$_{15}$~network, where each~$a_i$ has place value~$i$ (\S~\ref{sec:csareduction}).
\item Initialise 16-bit counters vectors $c=(c_0, c_1, \ldots, c_{2w_{\max}/r})$ to zero.
\item Until less than $16r$ bytes of input remain:
\begin{enumerate}
\item Read 16~vectors $v_0, v_1, \ldots, v_{15}$ from the input array.  Reduce
them with accumulators~$(a_8, a_4, a_2, a_1)$ into $(a_{16}, a_8, a_4, a_2, a_1)$ using
another CSA$_{16+4}$~network (\S~\ref{sec:csareduction}).  Keep $(a_8, a_4, a_2, a_1)$ for the next iteration.
\item Deinterleave and reduce the bits in $a_{16}$ and add them to the counter vectors (\S~\ref{sec:intermedaccum}).
\item If the counter vectors could overflow in the next iteration or during postprocessing,
call $f_w(C,c)$ to flush the counter vectors into the counter array, then reset $c$ to zero (\S~\ref{sec:accumoverflow}).
\end{enumerate}
\item With less than $16r$ bytes of input remaining, transpose and reduce
$(a_8, a_4, a_2, a_1)$ and add the counts to~$c$ (\S~\ref{sec:final-accumulation}).
\item Process the remaining bytes of input using a special algorithm for short inputs and
add their counts to~$c$ (\S~\ref{sec:short-arrays-scalar-tail}).
\item Call $f_w(C,c)$ to add the remaining counts to the counter array.
\end{enumerate}
\noindent

\begin{figure}
\centering
\subfigure[Algorithm sketch for Harley-Seal]{
\includegraphics[width=0.4\textwidth]{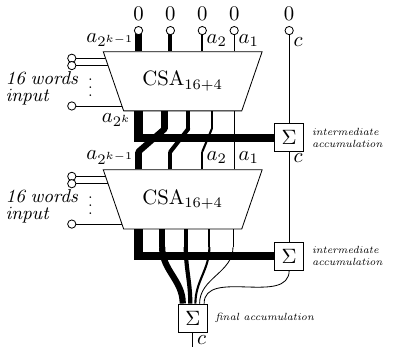}
\label{fig-alg-harley-seal}}\quad
\subfigure[Algorithm sketch with separate CSA$_{15}$ network for the initial 15 words]{
\includegraphics[width=0.4\textwidth]{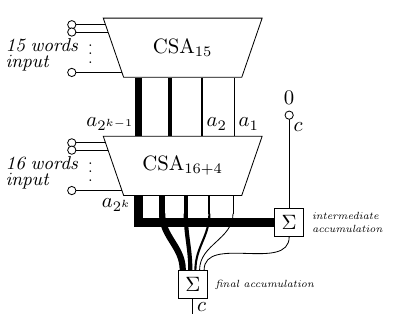}
\label{fig-alg-clausecker}}
\caption{Algorithm sketches with carry save adder (CSA) networks
\label{fig:alg-csa-networks}}
\end{figure}

This algorithm employs a structure similar to the Harley-Seal algorithm, but differs in that instead of
initialising the accumulators to zero, we run an initial iteration of only 15~vectors of input
with no accumulators carried in (cf.~Fig.~\ref{fig:alg-csa-networks}).
While processing one vector of input less, this initial
CSA~network requires only 11~full adders instead of the 15~adders used in the main loop, leading
to an overall time save.  For comparison, Alg.~\ref{alg:ours} shows the scheme used for our
algorithm in a manner similar to Alg.~\ref{alg:harleyseal} and Alg.~\ref{alg:klarqvist}.

\begin{algorithm}
\caption{Simplified algorithm sketch without vectorization, head/tail processing, and $f_w(C,c)$} \label{alg:ours}
\begin{algorithmic}
\Require{\textrm{$A$ is an array of $n\times2^k-1$~words of $w$~bits each}}
\State $(a_{2^{k-1}}, \ldots, a_2, a_1)\leftarrow\textrm{CSA}(A[0],A[1],\ldots,A[2^k-2])$ \Comment{\textrm{\S~\ref{sec:csareduction}}}
\State $c[0,1,\ldots,w_{\max}-1]\leftarrow0$
\For{$i\leftarrow0,2^k,\ldots,(n-2)2^k$}
\State $a_{2^k},(a_{2^{k-1}},\ldots,a_2,a_1)\leftarrow\textrm{CSA}(a_{2^{k-1}},\ldots,a_2,a_1;\;A[2^k+i-1],A[2^k+i],\ldots,A[2\cdot2^k+i-2])$ \Comment{\textrm{\S~\ref{sec:csareduction}}}
\State Accumulate $a_{2^k}$ into $c$ \Comment{\textrm{\S~\ref{sec:intermedaccum}}}
\EndFor
\State Accumulate $(a_{2^{k-1}},\ldots,a_2,a_1)$ into $c$ \Comment{\textrm{\S~\ref{sec:final-accumulation}}}
\State \Return $c$
\end{algorithmic}
\end{algorithm}

\subsection{Head processing} \label{sec:headprocessing}
On many microarchitectures, data is processed faster if it is accessed from aligned
addresses.  On some, this is a hard requirement (unaligned access faults).  To address
this need, we start out by aligning the input buffer to a multiple of $r/8$~bytes.  The
address is rounded down to the previous multiple of~$r/8$ and an initial vector~$v_0$ is
loaded from the aligned address.  This load cannot fault as the page size is a multiple
of~$r/8$ and thus a load from an aligned vector cannot cross a page.  All bits in this
vector that preceed the beginning of the vector are cleared, removing any influence of
surrounding data on the result of the algorithm.
The $8$ to~$r$ data bits in~$v_0$ are then complemented by additional 14~vectors of input
$v_1, v_2,~\ldots, v_{14}$ to serve as input for the initial CSA~reduction.
This approach is an improvement over the Klarqvist~et~al.\ method which aligns the input
by processing the first up to $r/8$~bytes of input in a scalar manner until sufficient
alignment is reached.

\subsection{Reduction using CSA networks} \label{sec:csareduction}
Like in the Klarqvist\ et\ al.~algorithm, each chunk of input is initially reduced using
a network of carry-save adders.  This eliminates the need for any further processing for
most input bytes, leaving us with just one vector of data to be processed in the following
steps.

We use two CSA~networks (cf.~Fig.~\ref{fig:csa-networks}): CSA$_{15}$ turns the initial
15~vectors of input $v_0, v_1, \ldots, v_{14}$ into vectors $(a_8, a_4, a_2, a_1)$ forming
a 4-bit accumulator, and CSA$_{16+4}$ adds 16~vectors of input $v_0, v_1, \ldots, v_{15}$ to
the same $(a_8, a_4, a_2, a_1)$, yielding a 5-bit accumulator $(a_{16}, a_8, a_4, a_2, a_1)$.
The top vector of bits~$a_{16}$ is then skimmed off and processed, leaving the other vectors
for the next iteration.

The specific CSA~networks we use have been designed for good instruction-level parallelism
(ILP) and are constrained largely by register pressure.  A variety of CSA~network designs are
possible and there is likely a different optimal design for each ISA\@.  The authors have not
exhaustively explored all possible CSA~networks, but have instead chosen one empirically.

In contrast to the Klarqvist\ et\ al.~algorithm, we operate with a fixed accumulator of 4~bits:
the final accumulation step involves a sequence of transposition steps that work best on a number of
vectors that is a power of two.  From the benchmarks of the Klarqvist\ et\ al.~algorithm, it
is clear that using only a 2-bit accumulator (corresponding to their \SI{256}{\byte} block
size) yields poor performance, while an 8-bit accumulator would only bring benefits for
unreasonably large inputs.

Another difference is that the Klarqvist\ et\ al.~algorithm does not use a separate
CSA~network for the initial chunk of input.  Instead, it initialises the accumulator
vectors to zeroes and starts directly with the main loop.  This is suboptimal for two
reasons: (a)~while the main loop CSA~network processes one more vector of input bytes,
it takes 4 more CSA steps to do so, a disproportionate amount of extra work.  (b)~with
a dedicated initial iteration, no~$a_{16}$ vector is produced for the initial $15r$~bytes
of input, avoiding the need to reduce that vector into the counters and saving time.

\begin{figure}
\centering
\includegraphics[width=.7\textwidth]{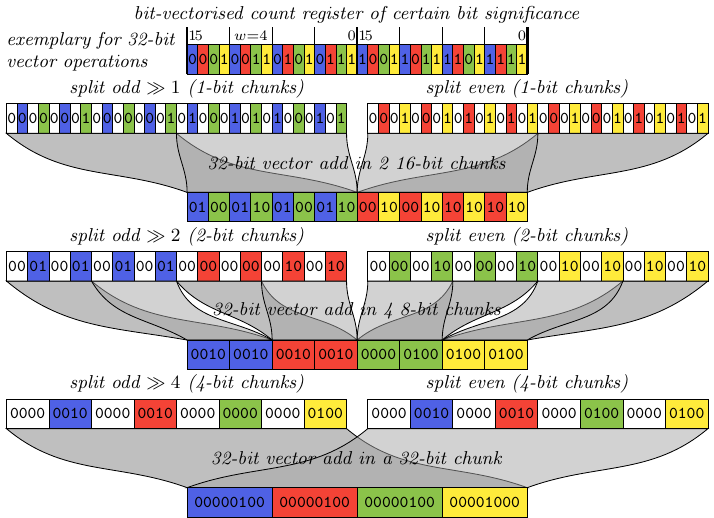}
\caption{Intermediate, vectorised accumulation procedure (see \S~\ref{sec:intermedaccum})
\label{fig:accu-intermediate}}
\end{figure}

\subsection{Bit accumulation: main loop}
\label{sec:intermedaccum}
After having reduced the input $v_0, v_1, \ldots, v_{15}$ with $(a_8, a_4, a_2, a_1)$ into
$(a_{16}, a_8, a_4, a_2, a_1)$, we keep $(a_8, a_4, a_2, a_1)$ for the next iteration and
add the bits in $a_{16}$ to the accumulators~$c$.  To do so, each bit of~$a_{16}$ needs to be
zero-extended into a 16-bit vector element and the vector has to be folded over itself until
no more than $w_{\max}$~elements remain.  Then, the result is scaled by~16 and added to~$c$.
I.\,e. we want to compute
\begin{equation}\label{mainaccum}
c[i]_{16} \leftarrow c[i]_{16} + 16 \sum_{j=0}^{r/w_{\max}-1} a_{16}[\/jw_{\max}+i\/]_1.
\end{equation}
This step is also required in the Klarqvist~et~al.\ algorithm, but while they implemented
Eq.~\ref{mainaccum} by looping over~$c[i]$, we provide an improved, fully vectorised approach.

The bits of~$a_{16}$ are repeatedly split into even and odd bits and folded over themselves,
increasing the size of each element from bits to crumbs to nibbles to bytes, possibly up to
16-bit words, while halving the number of elements in each step.  This is repeated
until~$w_{\max}$ elements remain or the element size has reached 16~bits, whatever happens
first.  The result is then zero-extended to 16-bit elements using the same deinterleaving steps,
but without the subsequent reduction, or using dedicated zero-extension instructions, and added
to~$c$.

The exact steps and shuffles needed to implement this reduction depend on~$r$ and~$w_{\max}$ and
cannot be given for the general case.  A synthetic example for $r=32$, $w_{\max}=4$ is given in
Fig.~\ref{fig:accu-intermediate}.  As can be seen, each iteration requires a different permutation
schedule.  This corresponds roughly to the start of our AVX-512 implementation's part that
accumulates $a_{16}$ into~$c$, using $r=512$, $w_{\max}=64$:

\begin{lstlisting}[style=customc,language={[Motorola68k]Assembler},morecomment={[l]{\#}},morecomment={[l]{;}},morekeywords={vpandd,vpandnd,vpslrd,vpsrld,vshufi64x2,vpaddd,vpslld}]
; convert zmm4 = a16 to bytes and reduce thrice
; zmm28 = 0x5555..55, zmm27 = 0x3333..33, zmm26 = 0x0f0f..0f
vpandd     zmm5, zmm28, zmm4         ; zmm4 &  0x5555..55  (bits 02468ace x32)
vpandnd    zmm6, zmm28, zmm4         ; zmm4 & ~0x5555..55  (bits 13579bdf x32)
vpsrld     zmm6, zmm6, 1             ; zmm6 shifted to the right by 1
vshufi64x2 zmm10, zmm5, zmm6, 0x44   ; zmm10 = 02468ace x16 13579bdf x16 (low)
vshufi64x2 zmm11, zmm5, zmm6, 0xee   ; zmm11 = 02568ace x16 13579bdf x16 (high)
vpaddd     zmm4, zmm10, zmm11        ; zmm4 = zmm10 + zmm11 (first reduction)

vpandd     zmm5, zmm27, zmm4         ; zmm4 &  0x3333..33  (048c x16  159d x16)
vpandnd    zmm6, zmm27, zmm4         ; zmm4 & ~0x3333..33  (26ae x16  37bf x16)
vpsrld     zmm6, zmm6, 2             ; zmm6 shifted to the right by 2
vshufi64x2 zmm10, zmm5, zmm6, 0x88   ; zmm10 = 048c x8 159d x8 26ae x8 37bf x8 (low)
vshufi64x2 zmm11, zmm5, zmm6, 0xdd   ; zmm11 = 048c x8 159d x8 26ae x8 37bf x8 (high)
vpaddd     zmm4, zmm10, zmm11        ; zmm4 = zmm10 + zmm11 (second reduction)

vpandd     zmm5, zmm26, zmm4         ; zmm4 &  0x0f0f..0f  (08 x8 19 x8 2a x8 3b x8)
vpandnd    zmm6, zmm26, zmm4         ; zmm4 & ~0x0f0f..0f  (4c x8 5d x8 6e x8 7f x8)
vpslld     zmm5, zmm5, 4             ; zmm5 shifted to the left by 4 (!)
vshufi64x2 zmm10, zmm5, zmm6, 0x88   ; zmm10 = 08 19 2a 3b 4c 5d 6e 7f (each x4, low)
vshufi64x2 zmm11, zmm5, zmm6, 0xdd   ; zmm11 = 08 19 2a 3b 4c 5d 6e 7f (each x4, high)
vpaddd     zmm4, zmm10, zmm11        ; zmm4 = zmm10 + zmm11 (third reduction)
\end{lstlisting}

The main differences are the different shuffles as well as the use of a left shift
in the last step.  This pre-scales the reduced counts by~$16$, removing the need for
an explicit multiplication with~$16$ prior to adding to~$c$.

\subsection{Accumulator overflow handling} \label{sec:accumoverflow}
In each iteration of the main loop, the elements of~$c$ are increased by at most
$16r/w_{\max}$.  Throughout the algorithm, we keep track of what the highest value~$h$
the elements of~$c$ could hold.  This value is initially~$h\leftarrow0$.  After each
iteration of the main loop, we increase $h$ by~$16r/w_{\max}$ and check if
\begin{equation}
h\le 2^{16} - 1 - (15+15)\frac{r}{w_{\max}}
\end{equation}
to ensure that there is enough space left before overflow occurs to fit at least one
more iteration of the main loop (up to~$16r/w_{\max}$) or the final accumulation
($15r/w_{\max}$) plus the maximum tail size (another~$15r/w_{\max}$).  If this is the
case, we proceed with the next iteration.  Otherwise we first call~$f_w(C,c)$ to flush
the counter vectors~$c$ into the counter array~$C$ and reset both $h$ and~$c$ back to
zero.

The function~$f_w(C,c)$ is implemented in much the same way as the intermediate accumulation
procedure from \S~\ref{sec:intermedaccum}: the counters in~$c$ are extended to the size
of the counters in~$C$ while being folded over until only $w$~counters remain.  Then the
resulting vectors are added to~$C$.

\subsection{Bit accumulation: final accumulation}
\label{sec:final-accumulation}
\begin{figure}
\centering
\includegraphics[width=.7\textwidth]{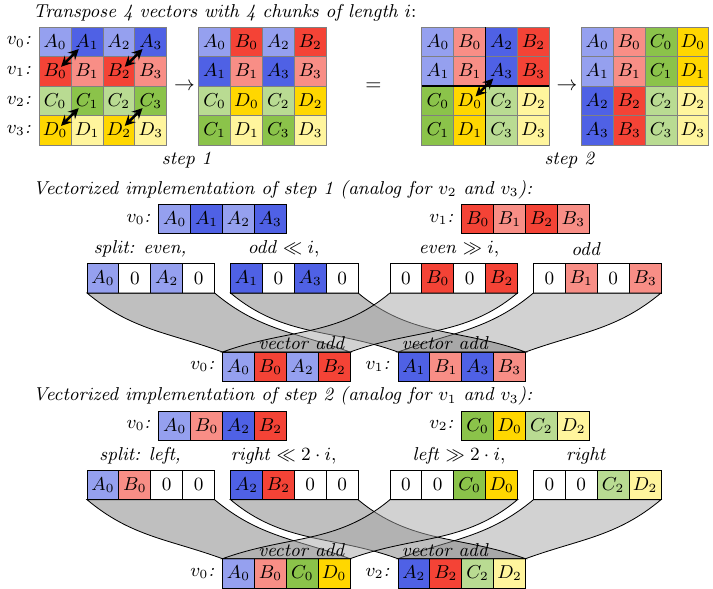}
\caption{Vectorised data restructuring (transposition) in the final accumulation procedure (see \S~\ref{sec:final-accumulation})}
\label{fig:transpose}
\end{figure}

Once less than $16r$~bits of input remain, the main loop is terminated and a final
accumulation step adds the contents of $(a_8, a_4, a_2, a_1)$ to~$c$.  This is done
in a similar way to the main loop accumulation procedure, except we start by turning
the four vectors of bits into one group of vectors, each of which holds a set of
nibble-sized counters.
This first step can be seen as a kind of transposition where $(a_8, a_4,
a_2, a_1)$ form a sequence of $4\times4$~matrices to be transposed.

The transposition is easily afforded using the classic recursive matrix transposition
algorithm\cite{Frigo1999}.  In this algorithm, a square matrix $M=\binom{A\;B}{C\;D}$ is split into
four equally sized block-matrices $A$, $B$, $C$, and $D$, which are recursively
transposed.  Finally, the top right and bottom left block matrices are swapped to give
the transposed matrix
$M^{\mathrm T}=\binom{A^{\mathrm T}\;C^{\mathrm T}}{B^{\mathrm T}\;D^{\mathrm T}}$.
We implement this transposition in a bit-parallel manner, as depicted in
Fig.~\ref{fig:transpose}.  In the AVX-512 implementation, this is realised as follows:

\begin{lstlisting}[style=customc,language={[Motorola68k]Assembler},morecomment={[l]{\#}},morecomment={[l]{;}},morekeywords={vpand,vpxor,vpor,bit,vmovdqa64,vpternlogd,vpsrld,vpslld}]
; transpose zmm3:zmm2:zmm1:zmm0 (a8:a4:a2:a1) into 4 vectors of nibbles
; zmm28 = 0x5555..55, zmm27 = 0x3333..33

vpsrld     zmm4, zmm0, 1            ; zmm4 = a1 >> 1
vpslld     zmm5, zmm1, 1            ; zmm5 = a2 << 1
vpsrld     zmm6, zmm2, 1            ; zmm6 = a4 >> 1
vpslld     zmm7, zmm3, 1            ; zmm7 = a8 << 1
vpternlogd zmm0, zmm5, zmm28, 0xe4  ; a12l = a1 & 0x55..5 | (a2 << 1) & 0xaa..a
vpternlogd zmm1, zmm4, zmm28, 0xd8  ; a12h = a2 & 0xaa..a | (a1 >> 1) & 0x55..5
vpternlogd zmm2, zmm7, zmm28, 0xe4  ; a48l = a4 & 0x55..5 | (a8 << 1) & 0xaa..a
vpternlogd zmm3, zmm6, zmm28, 0xd8  ; a48h = a8 & 0xaa..a | (a4 >> 1) & 0x55..5

// second step
vpsrld     zmm4, zmm0, 2            ; zmm4 = b12a >> 2
vpsrld     zmm6, zmm1, 2            ; zmm6 = b12b >> 2
vpslld     zmm5, zmm2, 2            ; zmm5 = b48a << 2
vpslld     zmm7, zmm3, 2            ; zmm7 = b48b << 2
vpternlogd zmm2, zmm4, zmm27, 0xd8  ; a_c = b48a & 0xcc..c | (b12a >> 2) & 0x33..3
vpternlogd zmm3, zmm6, zmm27, 0xd8  ; a_d = b48b & 0xcc..c | (b12b >> 2) & 0x33..3
vpternlogd zmm0, zmm5, zmm27, 0xe4  ; a_a = b12a & 0x33..3 | (b48a << 2) & 0xcc..c
vpternlogd zmm1, zmm7, zmm27, 0xe4  ; a_b = b12b & 0x33..3 | (b48a << 2) & 0xcc..c
\end{lstlisting}

This leaves us with four vectors $(a_a, a_b, a_c, a_d)$ holding the counts of
$(a_8, a_4, a_2, a_1)$ transposed into nibbles.  Unfortunately, the transposition
procedure leaves us with permuted elements.  $a_a$~holds counts for bits~$0, 4, 8, \ldots, r-4$,
$a_b$~for $1, 5, 9, \ldots, r-3$, $a_c$~for $2, 6, 10, \ldots, r-2$, and $a_c$~for $3, 7, 11,
\ldots, r-1$.

Afterwards, the contents of the vectors are zero extended first to bytes, then to
16-bit words.  Between zero extensions, we permute the vectors to restore the order of
elements and reduce the number of vectors until there is only one element for each
of the $w_{\max}$~bit positions using ideas analogous to those in \S~\ref{sec:intermedaccum}.

\subsection{Short Arrays and Scalar Tail}
\label{sec:short-arrays-scalar-tail}

At the end of the main loop, up to $16r-1$~bits of input may remain to be processed.
As these are too few bits to process using the procedure from \S~\ref{sec:intermedaccum},
a special tail handling algorithm is used to process the remaining input,
$w_{\max}$~bits\footnote{or $r/8$~bits, whichever is larger} at a time: a vector of
$w_{\max}$ byte-sized counters is prepared.  For each group of $w_{\max}$~bits of input,
those vector elements for which the corresponding bits are set are incremented.

For SIMD~extensions with \emph{predicate masks} such as AVX-512 and SVE, this is easily
achieved by preparing a vector of ones, loading $w_{\max}$~bits of input into a predicate
mask and then performing an addition with merge-masking of the ones to the counter
vector.\footnote{Equivalently, a vector of all~$-1$ may be subtracted.}
For SIMD~extensions without this feature, the procedure is more involved.  First, each
byte of input is replicated eight times.  These bytes are then masked with
\texttt{0x8040201008040201} to isolate each bit into its own byte.  The bytes are
compared with zero to obtain a value of $-1$ for “bit set” or $0$~for “bit clear.”  This
value is then subtracted from the counter vector, emulating the effect of a masked
subtraction:
\begin{lstlisting}[style=customc,language={[Motorola68k]Assembler},morecomment={[l]{\#}},morecomment={[l]{;}},morekeywords={vpand,vpbroadcastq,vpshufb,vpcmpeqb,vpsubb}]
; count 8 bytes from xmm6 into ymm0 and ymm1
; ymm3 and ymm7 hold suitable permutation masks
; ymm2 holds 8040201008040201 x4
vpbroadcastq ymm4, xmm6        ; ymm4 has bytes 7654:3210 x4
vpshufb      ymm5, ymm4, ymm7  ; ymm5 has bytes 7777:7777:6666:6666:...:4444:4444
vpshufb      ymm4, ymm4, ymm3  ; ymm4 has bytes 3333:3333:2222:2222:...:0000:0000
vpand        ymm5, ymm5, ymm2  ; mask out one bit in each copy of the bytes
vpand        ymm4, ymm4, ymm2
vpcmpeqb     ymm5, ymm5, ymm2  ; set bytes to -1 if bits were set
vpcmpeqb     ymm4, ymm4, ymm2  ; or to 0 otherwise
vpsubb       ymm1, ymm1, ymm5  ; add 1/0 (subtract -1/0) to/from counters
vpsubb       ymm0, ymm0, ymm4
\end{lstlisting}

This is repeated until the entire tail is consumed.  If sufficient registers are
available, multiple iterations of the loop can be interleaved.  Once less than
$w_{\max}$~bits of input remain, a final load of $w_{\max}$~bits is
performed and the bytes beyond the end of the array are masked out.  As we have
aligned our input to a multiple of~$w_{\max}$ early on in the procedure and as
$w_{\max}$~is a fraction of the page size, this load never crosses a page boundary
and thus cannot fault, even though it reads past the end of the array.  The final
bits of input are then processed as above.  The byte-sized counter vector is
added to~$c$ and $f_w(C,c)$ is called one final time to wrap up the procedure.

A similar approach is used for arrays that are shorter than $15r$~bits in total.
However, we don't initially align the input to a multiple of $w_{\max}$~bytes to
reduce overhead for very short arrays.  This causes some extra complications in
the final iteration, as loads past the end of the array may indeed cross into
unmapped pages if the input array was not aligned.  This is dealt with either
using fault-suppressing masked loads (AVX-512, SVE) or with an extra case distinction
and post processing.

\subsection{Discussion}
Through the development of this algorithm, we produced implementations for architectures
IA-32 (SSE, AVX2), Intel~64 (SSE, AVX2, AVX-512), as well as AArch64 (ASIMD) with good
results.  Due to the different constraints provided by each combination of instruction set
architecture and SIMD~extension, slight variations in the design and implementation of the
various implementations are present:

\begin{itemize}
\item As shown in \S~\ref{sec:csaimpl}, different full adder circuits are used depending
  on the bitwise operations provided by the SIMD~extension used.
\item Due to variations in register pressure between SIMD extensions and architecture,
  ranging from just 8~registers on IA-32 with SSE and AVX2 to 32~registers on ASIMD and
  AVX-512, some variation in the order of full adders used in the CSA~tree is present.
\item Large variations in the specific accumulation schedules exist due to differences
  in vector size (and hence the number of permutation/reduction steps needed to get down
  to~$w_{\max}=64$) as well in shuffle instructions available.  While SSE, AVX2, and
  AVX-512 provide very similar shuffles, the set provided by ASIMD is very different,
  resulting in large variations in the accumulation schedules, though at a similar overall
  number of instructions per vector of data.
\item The variation in available shuffles also affects head processing
  (see \S~\ref{sec:headprocessing}) and
  scalar tail (see \S~\ref{sec:short-arrays-scalar-tail}).  Masked operations
  provided by AVX-512 greatly simplify both the load of data blocks smaller than the
  length of a vector, and the conditional increment used to count the tail's population.
\item The IA-32 implementations are based on an earlier prototype of the algorithm,
  where instead of processing 16~vectors per iteration and carrying over $(a_8,a_4,a_2,a_1)$
  between iterations, each iteration processes 15~vectors worth of data into four
  vectors of ouput, which are then transposed into the counter vectors using what is
  now called the “final accumulation” procedure from \S~\ref{sec:final-accumulation},
  keeping no bit vectors between iterations.  Due to declining interest in the IA-32
  architecture, the code was not redesigned to integrate a Harley-Seal-like schedule when
  that was found to yield superior results.
\end{itemize}


\subsubsection{Variable-length Vectors}

Recently, variable-length vector extensions such as AArch64/SVE and RISC-V/RVV have
started to emerge.  Conceptually similar to SIMD instruction set extensions, variable-length
vector extensions are characterised by their vector length being a microarchitectural
parameter that varies from one implementation of the architecture to another.  For
example, while the Fujitsu A64FX processor implements the SVE vector extensions to AArch64
at a length of 2048~bits per register, the ARM Neoverse V1 processor implements SVE with
registers of only 128~bits.  Programs written for these extensions must be designed to
cope with whatever vector length the hardware provides, either by writing code agnostic
to vector length, or by providing a family of implementations for the various possible
lengths.

There was specific interest in evaluating these extensions for our algorithm.  An implementation
was attempted for AArch64/SVE, but problems quickly became apparent: while CSA~schedule
and head~processing are very straightforward to implement, it is not clear to
the authors how the intermediate and final accumulation steps can be carried out
effectively.  If fixed~$w_{\max}$ and data size for the counters~$c$ is used, the number
of vectors needed to store these at native vector length and the accumulation schedules
to transpose and reduce~$a_{16}$ resp.~$(a_8,a_4,a_2,a_1)$ into that size data varies
depending on the native vector length.

Two approaches obtain:
(a)~different intermediate and final accumulation procedures as well as implementations
of~$f_w(C,c)$ for each supported~$w$ are provided for each of the 5~possible native
vector lengths from 128 to 2048~bits.
This is both tedious to program and hard to test, as vector lengths longer than the
native vector length cannot effectively be tested without emulation.
(b)~the transposition/reduction schedule in the accumulation procedures treats the
accumulators and counters as if they were 128~bits long, but possibly processes multiple
128-bit chunks per vector at the same time, only reducing to one chunk in~$f_w(C,c)$.
While less code is required than for approach~(a), SVE only provides horizontal
reductions reducing directly to scalars, and thus seemingly still requires the programmer
to provide a different code path for each vector width.

In addition to this challenge, SVE lacks many important instructions available with ASIMD.
Most crucially, the \texttt{bsl}/\texttt{bit}/\texttt{bif} family of instructions required
for the faster full-adder circuit is absent, leading to a projected overall worse
performance than with ASIMD at the same or even double vector length.  Other missing
instructions include the zero-extending “DSP” arithmetic instructions as well as some of
the shuffles used as a part of the accumulation procedures.  The authors therefore decided
to defer work on an SVE implementation until these problems can be addressed.

\begin{table*}
\caption{Benchmarked algorithms}\label{tbl:kernels}
\begin{tabular*}{\textwidth}{@{\extracolsep\fill}lcccc@{}}\toprule
&\textbf{SIMD extension}&\textbf{vector length}&\textbf{max.\ word width}&\textbf{block size}\\
&&$r$&$w_{\max}$&$2^kr$\\
\midrule
\textbf{Clausecker et al.}&AVX-512&512~bits&64~bits&1024~bytes\\
\textbf{Clausecker et al.}&AVX2&256~bits&64~bits&\hphantom 0512~bytes\\
\textbf{Clausecker et al.}&ASIMD&128~bits&64~bits&\hphantom 0256~bytes\\
\midrule
\textbf{Klarqvist et al.}&AVX-512&512~bits&16~bits&1024~bytes\\
\textbf{Klarqvist et al.}&AVX-512&512~bits&16~bits&\hphantom 0512~bytes\\
\textbf{Klarqvist et al.}&AVX-512&512~bits&16~bits&\hphantom 0256~bytes\\
\midrule
\textbf{baseline}&---&---&16~bits&\hphantom{000}2~bytes\\
\textbf{roof\/line}&AVX2&256~bits&---&---\\
\textbf{roof\/line}&ASIMD&128~bits&---&---\\
\bottomrule
\end{tabular*}
\end{table*}

\section{Benchmarks}  \label{sec:evaluation}
For evaluation, we use implementations of our new algorithm in assembly
with ASIMD~($r=128$) for AArch64 and with AVX2 ($r=256$) and AVX512-F/BW ($r=512$)
for Intel~64. 
All implementations are based on a common kernel and a set of accumulation
functions $f_8(C,c)$, $f_{16}(C,c)$, $f_{32}(C,c)$, and~$f_{64}(C,c)$,
providing support for accumulation into word widths $w=8,16,32,64$.
We make our code
available for free\footnote{\url{https://github.com/clausecker/pospop},
\url{https://github.com/lemire/pospopcnt_avx512}}.

For benchmarks on Intel~64, these implementations are complemented by the original code of the Klarqvist
et~al.~algorithm (see \S~\ref{sec:pospopcount} and Alg.~\ref{alg:klarqvist}.
While the algorithm described in their paper\cite{Klarqvist2019} is generic,
the authors focus on use of AVX-512 with block sizes of 256, 512, or 1024~bytes,
corresponding to 4, 8, or 16~vectors of input to the CSA accumulation step.
Counts are then produced for a fixed word size of 16~bits.

We also included a
scalar implementation similar to Fig.~\ref{fig:pseudocpp}, but manually unrolled
for a word size of~$w=16$ and with auto-vectorisation inhibited, as well as
a dummy implementation that computes the sum of the input array and adds it to
to the output array.  Auto-vectorisation was enabled, letting the compiler
provide vectorised code for ASIMD and AVX2.\footnote{While the vector length is
shorter with AVX2 than it is with AVX-512, the same maximum memory bandwidth can
be achieved.}  This gives us both a baseline for the minimal performance
expected as well as a roofline\cite{Williams2009} for the maximal performance to
be expected given the required memory accesses.

\subsection{Benchmark Design}

To analyze the performance of our method, we wrote a benchmark
harness patterned after the benchmarking mechanism shipped with
the Go programming language.
This harness differs from the Klarqvist et~al.\
benchmark~\cite{Klarqvist2019} as we made different design choices to
obtain a more meaningful result.
A direct comparison with the Klarqvist et~al.\ algorithm is possible
as we have added their code to our new benchmark framework.

The benchmark harness executes a positional population count
kernel on an array of $n$~bytes for a number of times~$k$.  For faster
setup, the arrays are left containing all zeroes, as none of the
algorithms benchmarked depend on the specific values being processed.
With each round, the number of iterations is repeatedly increased
in a geometric progression until the benchmark reaches an execution
time of at least two seconds.
Even if the first round already exceeds the target execution time,
the benchmark is executed for at least two rounds with only the
benchmark results of the final run being taken into account for the
results, guaranteeing the presence of at least one second of warm
up.

Both the input and counter arrays are re-used between iterations,
introducing a desired loop-carried dependency between consecutive
benchmark iterations.
The counter array is finally summed into an accumulator variable
qualified as volatile to prevent deletion of the benchmarking code
through an overly zealous compiler.

For each benchmark round, we measure the real time elapsed~$t$, the
number of cycles elapsed~$c$, and the number of instructions~$i$ executed.
Neglecting the execution time for the benchmark harness itself and
combined with $n$ and~$k$, this permits us to derive the following
benchmark results for each kernel and each array size~$n$:

\begin{tabbing}
  \=\hbox to 10.5em{\textbf{\sffamily speed:}\hfill}\=
  \parbox[t]{10.5cm}{Number of input bytes the kernel processed per second: $nk/t$} \\[1mm]
  \>\textbf{\sffamily cycles per byte:} \>
  \parbox[t]{10.5cm}{Number of CPU cycles the kernel takes to process one input byte: $c/nk$} \\[1mm]
  \>\textbf{\sffamily instructions per byte:}  \>
  \parbox[t]{10.5cm}{Number of instructions that are issued to process one byte of input: $i/nk$} \\[1mm]
  \>\textbf{\sffamily instructions per cycle:} \>
  \parbox[t]{10.5cm}{Average number of instructions per cycle through the benchmark: $i/c$}
\end{tabbing}

\begin{table*}
\caption{Properties of benchmark systems.\label{tab:bench-systems}}
\begin{tabular*}{\textwidth}{@{\extracolsep\fill}lccc@{}}\toprule
                                                  & \textbf{AVX2}             & \textbf{AVX-512}              & \textbf{ASIMD}       \\ 
\midrule
    \textbf{CPU}                                  & Intel Xeon W2133          & $\leftarrow\rlap{$^{\tnote{\bf a}}$}$ & AWS Graviton 3, c7g.large \\ 
    \textbf{instruction-set architecture}         & Intel 64                  & $\leftarrow$                  & AArch64             \\
    \textbf{microarchitecture}                    & Skylake                   & $\leftarrow$                  & Neoverse V1         \\ 
    \textbf{number of cores (hyper cores)}        & \num{6} (\num{12})        & $\leftarrow$                  & \num{64}            \\ 
    \textbf{base frequency}                       & \SI{3.6}{\GHz}            & $\leftarrow$                  & \SI{2.6}{GHz}       \\ 
    \textbf{turbo boost frequency (vector unit)}  & \SI{3.9}{\GHz} (\SI{3.5}{\GHz}) & $\leftarrow$            & n/a                 \\ 
    \textbf{vector width}                         & \SI{32}{\byte}            & \SI{64}{\byte}                & \SI{16}{\byte}      \\ 
    \textbf{number of vector registers / core}    & \num{16}                  & \num{32}                      & 32                  \\ 
    \textbf{L1d size / core}                      & \SI{32}{\kibi\byte}       & $\leftarrow$                  & \SI{64}{\kibi\byte} \\ 
    \textbf{L2 size / core}                       & \SI{1}{\mebi\byte}        & $\leftarrow$                  & \SI{1}{\mebi\byte}  \\ 
    \textbf{L3 size / shared}                     & \SI{8.25}{\mebi\byte}     & $\leftarrow$                  & \SI{32}{\mebi\byte} \\ 
    \textbf{RAM model, frequency}                 & DDR4, \SI{3.2}{\GHz}      & $\leftarrow$                  & DDR5                \\ 
    \textbf{RAM size (n$\times$DIMM size)}        & \SI{32}{\gibi\byte} (\num{4}$\times$\SI{8}{\gibi\byte})  & $\leftarrow$ & \SI{3.72}{\gibi\byte} \\ 
\midrule
    \textbf{operating system}                     & Debian 11 (bullseye)      & $\leftarrow$                  & Ubuntu 22.04.4 LTS             \\ 
    \textbf{kernel version}                       & Linux 5.10.0              & $\leftarrow$                  & Linux 6.2.0             \\ 
\bottomrule
  \end{tabular*}
  \begin{tablenotes}
    \item[$^{\textrm{a}}$] A left arrow indicates that the same value as in the next left column applies here as well.
  \end{tablenotes}
\end{table*}

\subsection{Benchmark Parameters}
The benchmarks were run on a system with an Intel Xeon W-2133 (Skylake) processor (see Table~\ref{tab:bench-systems}).
Turbo Boost was enabled so everything ran with corresponding frequencies.
For the duration of the benchmark, the CPU governor was configured
to not reduce the CPU speed and the system was idle except for the
single benchmark task (single-threaded). The benchmark thread was not pinned to a particular CPU core,
as thread-pinning was not found to affect the results.

The performance on arrays sized $2^i$~bytes for~$i=1$ (\SI{2}{\byte}) to~$i=30$
(\SI{1}{\gibi\byte}) as well as $3\cdot 2^i$~bytes for~$i=0$ (\SI{3}{\byte}) to~$i=29$ (\SI{1.5}{\gibi\byte})
was measured, showing the impact of L1~cache, L2~cache, L3~cache,
and finally main memory bandwidth on the performance while also
demonstrating the refinements for small arrays.
For comparability with the Klarqvist et~al.\ kernels~\cite{Klarqvist2019}, a word size of
$w=16$ is used, though the word size used is largely irrelevant
to the performance of our method as we use the same kernel for each~$w$
with just an accumulation function pointer swapped out.

For each array size, the \emph{Clausecker~et~al.}\ and \emph{Klarqvist~et~al.}\ algorithms
(see~Tbl.~\ref{tbl:kernels}) were evaluated and compared with the scalar~\emph{baseline}
implementation as well as a \emph{roof\/line}~kernel to estimate the maximum throughput to
be expected given the available memory bandwidth.

We repeated these measurements on an AArch64~based AWS Graviton 3 c7g.large instance
to benchmark the performance of the ASIMD~implementation of our algorithm.  As the
Klarqvist~et~al.\ kernels are only available for AVX-512, we excluded them from this
benchmark set.

\subsection{Evaluation}

\begin{figure}
\centering
\includegraphics[width=\textwidth]{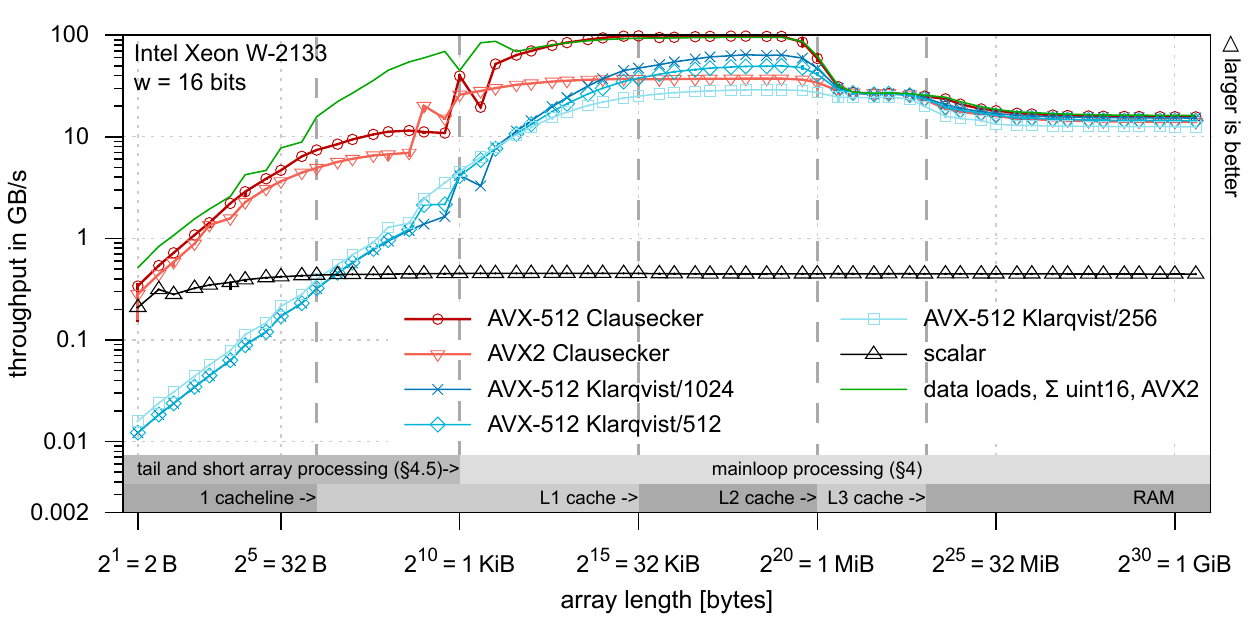}
\caption{AVX2 and AVX-512 speed of \pospopcnt{} as throughput per second by array length; mean of 10 at least 1 second runs
  per array size with minimum and maximum error bars (logarithmic
  scale on both axes).}
\end{figure}

The performance of our algorithm can roughly be modeled by three phases, depending on input length:
In the first phase, the special “short array” code from \S~\ref{sec:short-arrays-scalar-tail} is
used, processing 8~bytes of input per iteration.  Performance is initially dominated by the fixed
cost of calling $f_w(C,c)$~to add the results to the output array, but then converges to the
performance ceiling for the short array / scalar tail code.
Once the array size reaches $15r$~bits, the algorithm switches to the much faster CSA-based main algorithm,
leading to a sudden jump in performance.  The performance rises further, approaching the ceiling
given by memory bandwidth.  In the final stage, our array size exceeds the L2~cache, reducing the
maximum possible speed from the bandwidth of the L2~cache to that of main memory.

In contrast to the Klarqvist~et~al.\ algorithm, it can clearly be seen how performance matches
or exceeds the scalar implementation even for very short inputs.
A discontinuity is visible right after the CSA-based main algorithm is used for the first time.
This is caused by the next array size being $24r$~bits; large enough to process the initial
$15r$~bits using the CSA-based main algorithm, but just small enough to leave $9r$~bits to be
processed by the much slower scalar tail code.

\subsection{AVX-512}
Our AVX-512 implementation reaches its peak performance of~\SI{91.0}{\giga\byte\per\second} at an
input size of~\SI{512}{\kibi\byte}, a~$53\,\%$ improvement over the 1024~bit Klarqvist~et~al.\ kernel,
which clocks in at only~\SI{59.4}{\giga\byte\per\second} on the same input size.
This bandwidth is achieved while using only 0.09~instructions per byte of input, whereas
the 1024~bit Klarqvist~et~al.\ kernel requires 0.13~instructions per byte.

Notably, we actually slightly outperform the roofline reference (\SI{88.8}{\giga\byte\per\second}~at
the same input size), as it seems limited by a loop-carried dependency in summing the input data due
to poor auto-vectorised code generation.

\subsection{AVX2}
The AVX2 implementation reaches its peak performance of~\SI{34.8}{\giga\byte\per\second} at the
same input size as the AVX-512 kernel.  Unlike what one might naïvely expect based on the halved
vector length, the bandwidth is less than half of the AVX-512 kernel's bandwidth.  Various
factors affect this difference:

\begin{itemize}
\item While 512-bit SIMD instructions can be executed on two ports, three
 ports are available at a width of 256-bit as used with AVX2, leading to
 a higher maximum IPC.  This makes little difference in our case, as the
 instruction-level parallelism does not seem to be constrained by a lack
 of execution units, but rather by inherent limits of the algorithm.  The
 two in fact reach very similar IPC figures, with 2.54~IPC for the AVX-512
 kernel versus 2.49~IPC for the AVX2 kernel.
\item As explained in \S~\ref{sec:csaimpl}, the bit-parallel full adders
 used for the AVX2 kernel require 5~gates with a latency of 2/3~gates, whereas
 the AVX-512 kernel can make do with just a single gate for each of sum and
 carry, significantly increasing latency while reducing the opportunity for
 instruction-level parallelism.
\item This is slightly compensated by a faster intermediate accumulation
 procedure, as we need one less transposition step due to the shorter vector
 length.
\end{itemize}

\subsection{ASIMD}
Moving to AArch64, our ASIMD implementation plateaus at~\SI{16}{\giga\byte\per\second}
starting around a kilobyte of input data.  This is about $83\,\%$~of the
\SI{19.3}{\giga\byte\per\second}~roofline observed.  While much lower
than the AVX-512 performance, it appears fairly comparable when taking
into account the shorter vector length and the generally lower single-thread
performance of server processors as the Neoverse V1 we benchmarked on.

\begin{figure}
\centering
\includegraphics[width=\textwidth]{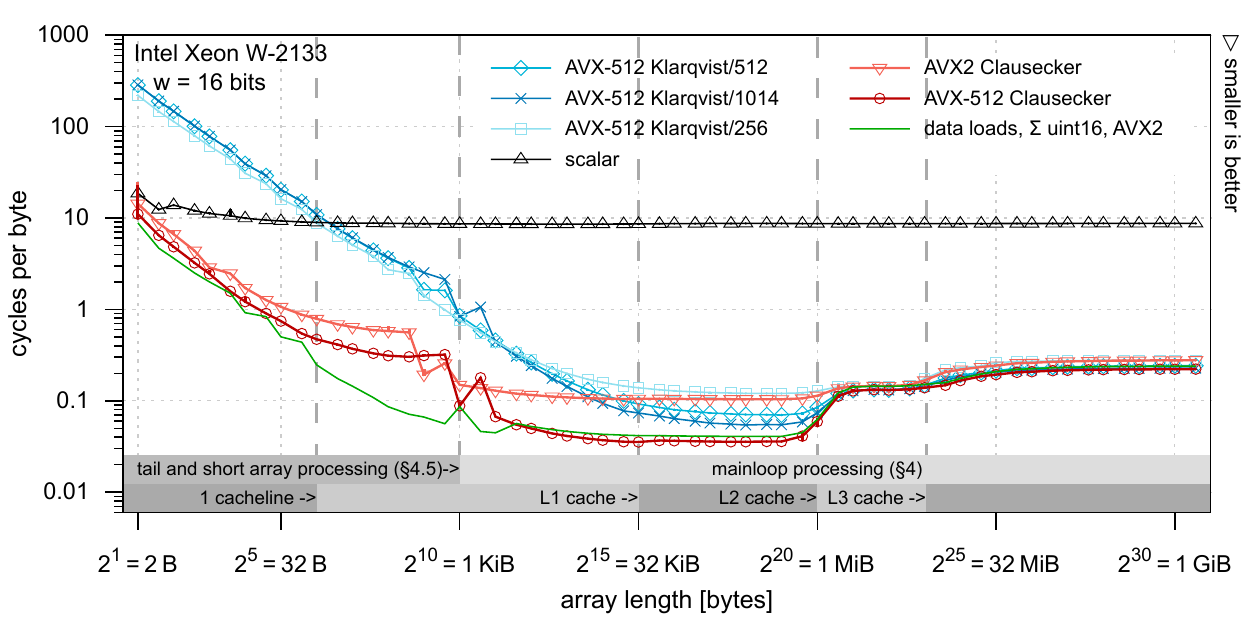}
\caption{AVX2 and AVX-512 cycles per byte of \pospopcnt{} by array length; mean of 10 at least 1 second runs
  per array size with minimum and maximum error bars (logarithmic
  scale on both axes).}
\end{figure}

\begin{figure}
\centering
\includegraphics[width=\textwidth]{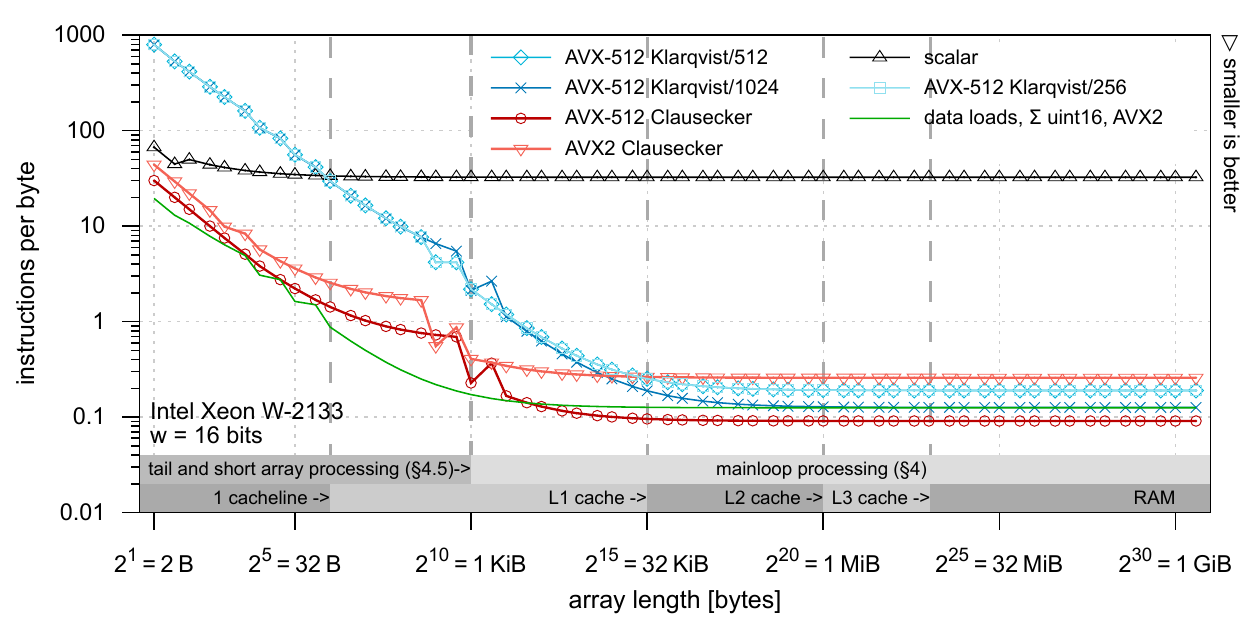}
\caption{AVX2 and AVX-512 instructions per byte of \pospopcnt{} by array length; mean of 10 at least 1 second runs
  per array size with minimum and maximum error bars (logarithmic
  scale on both axes).}
\end{figure}

\begin{figure}
\centering
\includegraphics[width=\textwidth]{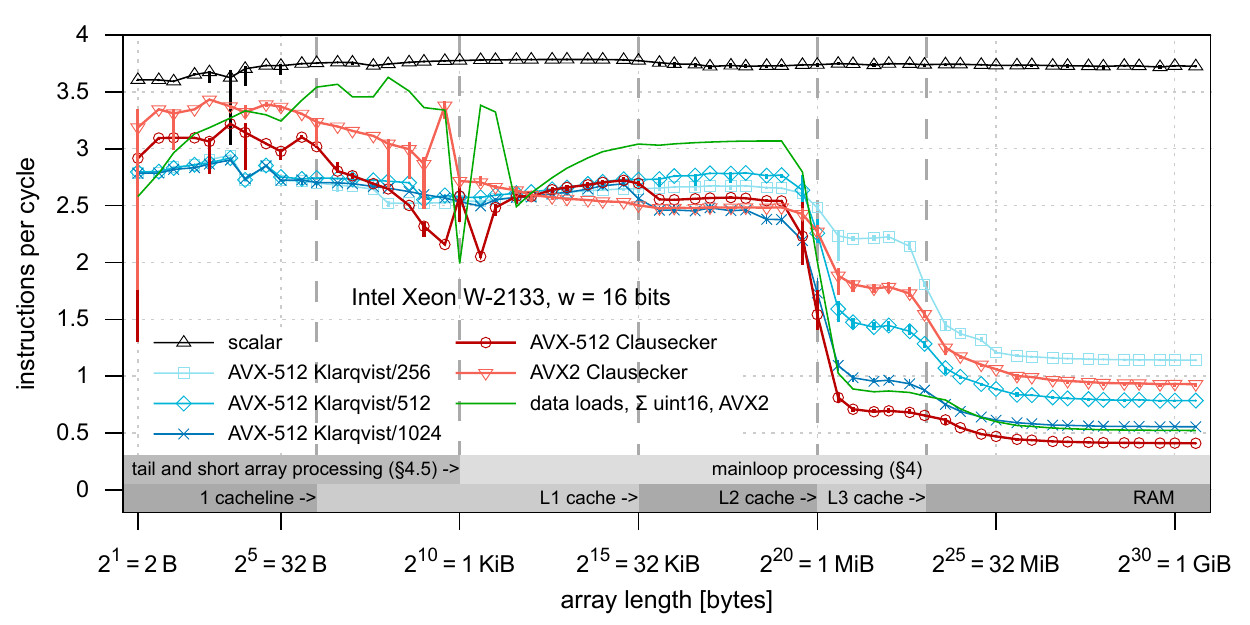}
\caption{AVX2 and AVX-512 instructions per cycle of \pospopcnt{} by array length; mean of 10 at least 1 second runs
  per array size with minimum and maximum error bars (logarithmic
  scale on the x-axis).}
\end{figure}

\begin{figure}
\centering
\includegraphics[width=\textwidth]{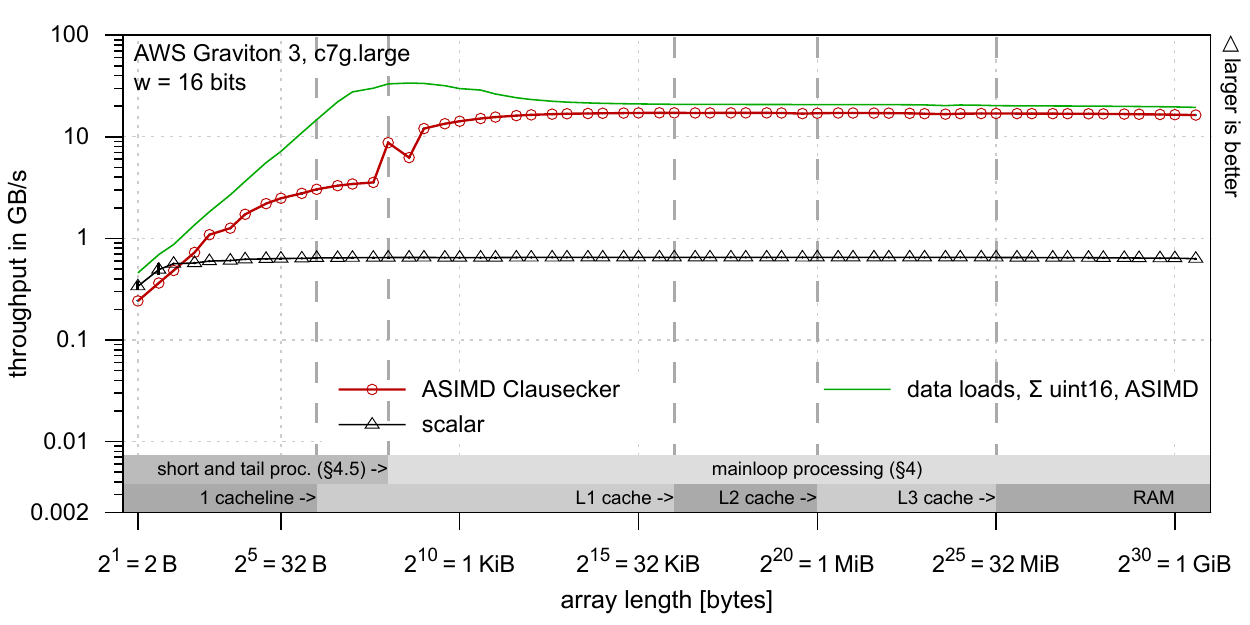}
\caption{ASIMD speed of \pospopcnt{} as throughput per second by array length; mean of 10 at least 1 second runs
  per array size with minimum and maximum error bars (logarithmic
  scale on both axes).}
\end{figure}

\section{Conclusion}
Using a modified Harley-Seal algorithm scheme coupled with bit-parallel
transposition/reduction logic, the positional population count operation can be computed efficiently,
with performance being memory bound for arrays larger than a few kilobytes.
The algorithm is implemented with a desired maximum word width~$w_{\max}$, allowing the user to
select any fraction of that word width as needed by program requirements at runtime.
Short arrays can be processed rapidly using a bit-parallel scheme, giving good performance for arrays
as small as~\SI{2}{\byte}.

Our approach is easily ported to new architectures:
Implementations for common SIMD instruction set extensions, including AVX2, AVX-512, and ASIMD, are provided
as open-source software.
Meanwhile, initial users of our improved algorithms are already found in the field of bioinformatics\cite{KMCP22}.

\section{Future Work}
In a 2024 blog post, Harold Aptroot proposes to use the Galois field affine transform
instruction \texttt{vgf2p8affineqb} available with AVX512-GFNI in conjunction with
cross-lane byte permutation instruction \texttt{vpermb} of AVX512-VBMI to regroup
the bits of eight 64-bit words within a 512-bit SIMD register such that each byte holds
the 8~bits of the same place value, permitting rapid computation of positional population
counts using the population count instruction \texttt{vpopcntb} available with
AVX512-BITALG\cite{Aptroot2024}.  While slower than our algorithm for long inputs,
its use for the acceleration of the intermediate accumulation procedure as well as
for the handling of short inputs and tail should be investigated.

\section*{Acknowledgements}
Robert Clausecker would like to thank Université TÉLUQ as well as the NHR-Verein Graduate School programme for
their research scholarships.

\bibliography{pospop}

\end{document}